\documentclass[aps,pre,twocolumn,showpacs,preprintnumbers,amsmath,amssymb]{revtex4-1}
\usepackage{epsfig}
\usepackage{graphicx}

\usepackage{color}
\usepackage{rotating}
\usepackage{amssymb}
\usepackage{amsmath}
\usepackage{physics}

\usepackage{dcolumn}% Align table columns on decimal point
\usepackage{bm}% bold math
\usepackage{epstopdf}

\def\beq{\begin{eqnarray}}
\def\eeq{\end{eqnarray}}

\newcommand{\be}{\begin{equation}}
\newcommand{\ee}{\end{equation}}
\newcommand{\bea}{\begin{eqnarray}}
\newcommand{\eea}{\end{eqnarray}}

\newtheorem{theorem}{Theorem}[section]

\begin{document}

\title{Thermal power of heat flow through a qubit}

\author{Erik Aurell}
\email{eaurell@kth.se}
\affiliation{
KTH -- Royal Institute of Technology,  AlbaNova University Center, SE-106 91~Stockholm, Sweden\\
Depts. Computer Science and Applied Physics, Aalto University, FIN-00076 Aalto, Finland \\
Laboratoire de Physico-Chimie Th\'eorique -- UMR CNRS Gulliver 7083, PSL Research University, ESPCI,
10 rue Vauquelin, F-75231 Paris, France
}

\author{Federica Montana}
\email{federica.montana@polito.it}
\affiliation{
Dept. Mathematics, Politecnico di Torino, Corso Duca degli Abruzzi, 24
10129 Torino, Italy\\
Nordita, Royal Institute of Technology and Stockholm University,
Roslagstullsbacken 23, SE-106 91 Stockholm, Sweden
}

\begin{abstract}
In this paper we consider thermal power of a heat flow through a qubit between two baths.
The baths are modeled as set of harmonic oscillators initially at equilibrium, at two temperatures.
Heat is defined as the change of energy of the cold bath,
and thermal power is defined as expected heat per unit time, in the long-time limit.
The qubit and the baths interact as in the spin-boson model, \textit{i.e.}
through qubit operator $\sigma_z$.
We compute thermal power in an approximation analogous to ``non-interacting blip'' (NIBA)
and express it in the polaron picture
as products of correlation functions of the two
baths, and a time derivative of a correlation function of the cold bath.
In the limit of weak interaction we recover known results
in terms of a sum of 
correlation functions of the two
baths, a correlation functions of the cold bath only,
and the energy split.
\end{abstract}
\pacs{03.65.Yz,05.70.Ln,05.40.-a}

\keywords{Stochastic thermodynamics, quantum power operators, quantum heat switches}
\maketitle

\section{Introduction}
\label{sec:introduction}
Heat and work in classical thermodynamics are properties of processes, 
and not states. 
Heat is further in classical thermodynamics 
energy transferred from the system to an uncontrolled 
environment such that it cannot later be retrieved to do useful 
work~\cite{Sekimoto-book,seifert12-review}.  
The translation of these concepts to the quantum domain is therefore not obvious, 
as discussed in an early review~\cite{Esposito09-review}.
Quantum thermal power is average quantum heat per unit time, and
is a centrally important topic for \textit{e.g.}
applications to quantum heat engines~\cite{KosloffLevy2014,Pekola2015,VinjanampathyAnders2016}.
While heat and thermal power at weak coupling 
has been studied for a long time in the literature~\cite{Weiss-book,Alicki-Lendi-book,KarimiPekola2016},
the attention to systems interacting strongly with one or
several baths is more recent, see cf~\cite{EspositoOchoaGalperin,Newman2017,GoyalKawai2017,Ronzani2018,Douetal2018,Kwonetal2018,PernauLlobet2018}.

The goal of this paper is to revisit these questions in perhaps the simplest 
non-trivial scenario: one qubit interacting strongly with two heat baths 
at different temperatures.
We will start from the general and unifying point of view that
heat is energy change in a bath. Thermal power is thus 
expected energy change in a bath per unit time, in the long-term limit.
For concreteness we will consider thermal power as heat per time to
the cold bath, and thus a quantity that has to be non-negative in the long term limit. 
We assume that the qubit interacts with the baths and with an external drive
as in the spin-boson model which allows to re-use many results 
developed in that literature~\cite{Leggett87}. 
At strong coupling, and in the approximation
known as ``non-interacting blip approximation'' (NIBA), 
the stationary state of the qubit is then
determined by equilibrium correlation functions of certain bath operators
related to a polaron transform. 
Our main result is that 
in a similar approximation thermal power is determined by
derivatives of the same correlation functions with respect to time. 

The paper is organized as follows: in Section~\ref{sec:model}
we introduce our model, and in Section~\ref{sec:arguments}
we give dimensional arguments what the results should be,
first in a version appropriate for weak coupling, and then in a version
appropriate for strong coupling.
Section~\ref{sec:strong-coupling} contains an overview
of the calculations, and states the results in path integral language
while Section~\ref{sec:polaron-transform} states in the language 
of the correlation functions after the polaron transform.
Section~\ref{sec:discussion} summarizes and discusses the results.

Some of the background and much of the calculations are presented in appendices.
Appendix~\ref{app:spin-boson} is thus a summary of the vast literature
on the spin-boson problem, sufficient for our purposes.
Appendix~\ref{sec:heat-and-NIBA} summarizes on the other hand earlier work
on quantum heat functionals~\cite{AurellEichhorn,Aurell2017,Aurell2018b}
adapted to the spin-boson setting, 
and Appendix~\ref{app:Ohmic-baths}
gives details of what these formulas mean for Ohmic baths.
Appendix~\ref{sec:bath-momentum-coupling} further 
translates this theory to when the interaction is through bath momentum.
Appendices~\ref{sec:singluar-heat-NIBA}-\ref{sec:long-term-limit-derivative}
finally contain details of the calculations presented in Section~\ref{sec:strong-coupling}.

\section{The model}
\label{sec:model}
We consider one qubit interacting 
with two harmonic oscillator baths as in the spin-boson model~\cite{Leggett87}.
Harmonic oscillator baths model, for instance, resistive elements in electrical
circuits, and quantum harmonic oscillator baths hence model how such elements interact
with other circuit elements at sufficiently low temperatures~\cite{Devoret1995}. 
Circuits with superconducting elements that can be assimilated to
qubits are widely investigated in  
scalable quantum information processing~\cite{Wendin2017}.
The state of one qubit interacting with two baths is hence a toy model
of a quantum computer perturbed by a heat flow
through the dynamical degrees of freedom of quantum computer itself.
Quantum thermal power in this setting is conversely how well such a device 
can transport energy between two baths in the quantum regime.   

The system, the baths and the interactions can thus be written down as a total Hamiltonian
\begin{equation}
H_{TOT} = H_S + H_C + H_H + H_{CS} + H_{HS} 
\label{eq:H_TOT}
\end{equation}
where ``$C$'' refers to the cold bath (temperature $T_C$) and ``$H$'' refers to the hot bath (temperature $T_H$).

The system Hamiltonian is 
\begin{equation}
H_S = -\hbar\frac{\Delta}{2} \hat{\sigma}_x +  \frac{\epsilon}{2} \hat{\sigma}_z 
\label{eq:H_S}
\end{equation}
where $\Delta$ is a rate (dimension $\left(\hbox{time}\right)^{-1}$), and $\epsilon$ is the level splitting.
The bath Hamiltonian are
\begin{eqnarray}
\label{eq:H_C}
H_C &=& \sum_{b\in C} \frac{p_b^2}{2m_b}+\frac{1}{2}m_b\omega_b^2 q_b^2 \\
\label{eq:H_H}
H_H &=& \sum_{b\in H} \frac{p_b^2}{2m_b}+\frac{1}{2}m_b\omega_b^2 q_b^2 
\end{eqnarray}
where the parameters $m_b$ and $\omega_b$ are the mass and angular frequency of each oscillator
and $C$ and $H$ also stand for the sets of oscillators in respectively the cold bath and the hot bath.

We will take the system-bath interactions to be described by
\begin{eqnarray}
\label{eq:H_LS}
H_{CS} &=& -\sum_{b\in C} C_b q_b \hat{\sigma}_z \\
\label{eq:H_RS}
H_{HS} &=& -\sum_{b\in H} C_b q_b \hat{\sigma}_z 
\end{eqnarray}
where $C_b$ is the interaction coefficient between bath oscillator $b$
and the qubit, $q_b$ is the oscillator coordinate, and $\hat{\sigma}_z$ operates on the qubit.
Pauli matrices are by convention dimension-less,
and the coupling coefficients $C_b$ hence have 
dimension $(\hbox{energy})\cdot (\hbox{length})^{-1}$.
In~\cite{Leggett87} the length scale (there called $q_0$) is taken to be
the spatial distance between the minima of two potential wells.
For a qubit formed out of a non-linear oscillator the length scale could
similarly be the typical spatial scale of the
oscillator ground state, $\sqrt{\frac{\hbar}{m\omega}}$.

We consider heat as related to two measurements on the cold bath, one at the beginning
of the process and one at the end, which we assume to take values 
${\cal E}_i$ and ${\cal E}_f$. In a quantum bath neither ${\cal E}_i$ nor ${\cal E}_f$
are known; all we can know is the probability of observing ${\cal E}_i$ 
at the beginning, and the probability of observing ${\cal E}_f$
at the end. Thermal power is then the expected change of 
bath energy per unit time $\frac{\left<{\cal E}_f-{\cal E}_i\right>}{t_f-t_i}$.

Four remarks are in order. First, ``measurement on the bath'' 
is required in the theory we consider, as without measurement the bath energy does not have a
definite value. However, expected heat per unit time 
can, as we will see, be expressed in terms of system properties alone.
Thermal power hence does not make 
any direct references to measurement, the values of which can hence be taken to be unrecorded.
We may thus imagine ``measurement on the bath'' to actually refer to
interaction with a large super-bath
which forces the bath states to decohere, without assuming any direct control of the bath states
by an experimenter.
Second, we do not count any part of the interaction energy in 
the heat. While this issue is important and has been discussed at length on the classical side
in the recent literature~\cite{Seifert2016,TalknerHanggi2016,Jarzynski2017,MillerAnders2017,Aurell2017,Aurell2018a},
it is reasonable to assume that the interaction energy between one qubit and a bath does not 
increase at a non-zero rate for long enough times. 
Third, in applications to superconducting circuits, the system-bath interaction may 
often more naturally be taken to be 
proportional to bath oscillator momentum variable $p_b$~\cite{Devoret1995}.
Since both $q_b$ and $p_b$ can be expressed in Fourier modes of
the oscillator this can be expected to make no essential
difference, as was indeed stated in~\cite{CaldeiraLeggett83a} for the qubit state.
For completeness we outline in Appendix~\ref{sec:bath-momentum-coupling}
an argument that this is so also for heat (full distribution function of bath energy change).
Lastly, in realistic mesoscopic devices effective temperatures 
of different parts may differ. Such situations fall outside
what is considered here, since the devices would then not 
be systems in thermal equilibrium that could be modelled as baths.

\section{Dimensional arguments}
\label{sec:arguments}
The long-time limit of the state of one qubit interacting with any number of baths 
is given by its density matrix, where the diagonal terms (``the populations'') determine the
probability for the qubit to be respectively in the up state and in the down state.
Suppose these probabilities are $P\left(\hbox{up}\right)$
and $P\left(\hbox{down}\right)$. 
Suppose further that the memory of the bath is short enough
that when the system is in one state the bath does not remember in which
states the system was before.
We can then suppose that the expected energy given to the cold bath per unit
time takes two values that depend on the system state, call them
$\pi_{\hbox{up}}$ and $\pi_{\hbox{down}}$.
Thermal power can then be estimated as
\begin{equation}
\label{eq:dimensional-argument}
\Pi = P\left(\hbox{up}\right) \pi_{\hbox{up}} +  P\left(\hbox{down}\right) \pi_{\hbox{down}}
\end{equation}
To turn this into a quantitative prediction we can suppose that qubit transitions happen 
with effective rates describing the interactions
with the two baths, and call these rates
$\Gamma^C_{\downarrow\uparrow}$, $\Gamma^H_{\downarrow\uparrow}$,
$\Gamma^C_{\uparrow\downarrow}$ and $\Gamma^H_{\uparrow\downarrow}$.
This approach is appropriate 
when the qubit is weakly coupled to the baths,
and one considers sufficiently long time scales~\cite{Weiss-book,Alicki-Lendi-book}.
The up and down probabilities then depend on the rates as for a classical jump process
\textit{i.e.} as
\begin{eqnarray}
P\left(\hbox{up}\right)&=&\frac{\Gamma^C_{\downarrow\uparrow}+\Gamma^H_{\downarrow\uparrow}}{\Gamma^C_{\downarrow\uparrow}+\Gamma^H_{\downarrow\uparrow}+\Gamma^C_{\uparrow\downarrow}+\Gamma^H_{\uparrow\downarrow}}\nonumber \\
P\left(\hbox{down}\right)&=&\frac{\Gamma^C_{\uparrow\downarrow}+\Gamma^H_{\uparrow\downarrow}}{\Gamma^C_{\downarrow\uparrow}+\Gamma^H_{\downarrow\uparrow}+\Gamma^C_{\uparrow\downarrow}+\Gamma^H_{\uparrow\downarrow}}\nonumber
\end{eqnarray}
Power is dimensionally energy per unit time. 
When interaction energy is negligible the characteristic scale of
energy transferred to the cold bath must be  
$\epsilon$ in an up-to-down transition,
and 
$-\epsilon$ in an down-to-up transition,
and these happen with rates $\Gamma^C_{\uparrow\downarrow}$ and $\Gamma^C_{\downarrow\uparrow}$.
This leads to the estimates of power in the two states as
\begin{eqnarray}
\label{eq:power-weak-coupling-up}
\pi_{\hbox{up}}    &=& \epsilon \Gamma^C_{\uparrow\downarrow}\\
\label{eq:power-weak-coupling-down}
\pi_{\hbox{down}}  &=& -\epsilon \Gamma^C_{\downarrow\uparrow}
\end{eqnarray}
and overall expected power as
\begin{eqnarray}
\label{eq:power-weak-coupling}
\Pi^{\hbox{weak}} &=& \epsilon \left(\Gamma^C_{\uparrow\downarrow} P\left(\hbox{up}\right) - \Gamma^C_{\downarrow\uparrow}P\left(\hbox{down}\right) \right)
\end{eqnarray}
Expressions of this form are well known in the literature, \textit{e.g.} in~\cite{KarimiPekola2016} (Eq.~5), and essentially hold in weak coupling also without the assumption of a short bath memory time.  

At strong coupling the above is however not correct
because when the qubit flips there is also a change of interaction energy between qubit and
the bath. When this is larger than the level splitting 
the characteristic scale of energy transferred to the bath 
can be very different from $\epsilon$.
Furthermore, in strong coupling one may assume 
combined effective mean switching rates $\Gamma_{\uparrow\downarrow}$
and $\Gamma_{\downarrow\uparrow}$, but it is  
not possible to disentangle the actions of the
two baths into separate terms $\Gamma^C$ and $\Gamma^H$. 

A different argument can nevertheless be made
using the assumption of short enough bath de-correlation time,
or equivalently that $\Delta$ is small enough that 
the residence time of the qubit in one state is long enough.
From one qubit jump to the next qubit jump the baths hence on the average behave as follows.
Right after the jump into state $s$ there will be some average interaction energy
and some average bath energy, $\left<H_{CS}^i(s)\right>$ and $\left<H_{C}^i(s)\right>$.
Between the jumps, when the qubit does not change its state, the sum of these energies is conserved,
but in the same time interval the baths will equilibriate with the qubit.
At the end of the interval the average interaction energy should 
hence vanish. This means that during one residence time in state $s$
the expected energy change of the bath should be the expected initial interaction energy
\textit{i.e.}
$\left<H_{CS}^i(s)\right>$. 
By this reasoning one gets 
\begin{eqnarray}
\label{eq:power-strong-coupling}
\Pi^{\hbox{strong}}  &=& P\left(\hbox{up}\right)\Gamma_{\uparrow\downarrow} \left<H_{CS}^i(\hbox{up})\right> \nonumber \\
&&+ P\left(\hbox{down}\right)\Gamma_{\downarrow\uparrow} \left<H_{CS}^i(\hbox{down})\right>
\end{eqnarray}
The main contribution of this paper is to derive an estimate like
\eqref{eq:power-strong-coupling} systematically, and 
explain how the terms follow from the microscopic parameters of the model.

\section{Thermal power at strong coupling}
\label{sec:strong-coupling}
We now describe an approach to thermal power
at strong coupling based on the Feynman-Vernon formalism~\cite{FeynmanVernon}.
To calculate heat (energy change in a bath) we follow~\cite{AurellEichhorn,Aurell2017,Aurell2018a},
related general results can also been found in~\cite{Carrega2015,Carrega2016} and~\cite{FunoQuan2018}.
Adapting the Feynman-Vernon formalism to describe the development
of one spin interacting with one bath (the spin-boson problem)
is already not trivial~\cite{Leggett87}.
Here we have the complications that we are interested in heat in a 
spin interacting with two (or more) baths at different temperatures.
Technical background and details have therefore been moved to appendices;
here we only outline the main idea of the calculation.

We focus on the energy changes of one bath, for concreteness we assume that is the cold bath.
The starting point is to assume that initially the baths are independently at thermal equilibrium (at different temperatures),
and the system as well as the energy of the cold bath are measured.
After that measurement the state of the system and the baths 
is $\rho_H^{eq}\oplus \dyad{{\cal E}_i^{(C)}, i}$ where  $\rho_H^{eq}$ is the equilibrium state of the hot bath (or baths)..
$i$ indicates the state of the system after measurement and ${\cal E}_i^{(C)}$ the state of the cold bath. 
We take $p_C(\Delta E, f|{\cal E}_i^{(C)}, i)$ to be the conditional probability of observing
a final state $\ket{f}$ of the system and energy change of the cold bath
$\Delta E$, conditioned on total initial state.

Next we assume that the measured energy of the cold bath is not recorded.
This means that we could also say that the cold bath de-coheres by interacting with an unobserved 
cold super-bath at the same temperature.
The initial state of the cold bath is then a statistical mixture where $\ket{{\cal E}_i^{(C)}}$
appears with the Gibbs weight
$Z_C^{-1}(\beta)\exp\left(-\beta E({\cal E}_i^{(C)})\right)$.
Here $\beta$ is the inverse temperature of the cold bath, and $Z_C$ is the partition function.
From here we consider the average distribution
\begin{equation}
\overline{p}_C(\Delta E, f|i)= \sum_{{\cal E}_i^{(C)}} p_C(\Delta E, f|{\cal E}_i^{(C)}, i)
\frac{e^{-\beta E({\cal E}_i^{(C)})}}{Z_C(\beta)}
\end{equation}
which can be re-written
\begin{eqnarray}
\overline{p}_C(\Delta E,f |i) &=& \sum_{{\cal E}_f,{\cal E}_i} Z_B^{-1}(\beta) e^{-\beta E({\cal E}_i)} \mathbf{1}_{E({\cal E}_f)-E({\cal E}_i),\Delta E} \nonumber \\
&&\quad \matrixel{{\cal E}_f,f}{\rho_{TOT}({\cal E}_i,i)}{{\cal E}_f,f}
\end{eqnarray}
where $\rho_{TOT}({\cal E}_i,i)$ is the total density operator of the system and the bath at the end 
of the process, when the system and the cold bath started in the pure state $\ket{{\cal E}_i, i}$.
Resolving the delta function one can write 
\begin{equation}
\overline{p}_L(\Delta E,f|i)=\frac{1}{2\pi} \int e^{-i\nu\Delta E} G_{if}(\nu) d\nu 
\end{equation}
where 
\begin{eqnarray}
G_{if}(\nu) &=& \sum_{{\cal E}_f,{\cal E}_i} Z_B^{-1}(\beta)e^{-\beta E({\cal E}_i)} 
e^{i\nu\left(E({\cal E}_f)-E({\cal E}_i)\right)} \nonumber \\
&&\quad \matrixel{{\cal E}_f,f}{\rho_{TOT}({\cal E}_i,i)}{{\cal E}_f,f}
\end{eqnarray}
By linearity the Gibbs weight and the factor $e^{-i\nu E({\cal E}_i)}$
can be taken inside the the big unitary transformation defining
$\rho_{TOT}({\cal E}_i,i)$. The above is therefore the same as
\begin{eqnarray}
\label{eq:FCS-main}
G_{if}(\nu) &=& \hbox{Tr}_{CH}\matrixel{f}{e^{i\nu H_{C}} \left(U e^{-i\nu H_{C}} \rho_i^{TOT}\right) U^{\dagger} }{ f}
\end{eqnarray}
where $\rho_i^{TOT} = \rho_H^{eq}\oplus \dyad{i}\oplus \rho_C^{eq}$,
and the trace is over the cold and the hot bath(s).

$G_{if}(\nu)$ codifies all the information on the distribution of
energy change in a bath (here the cold bath), averaged over an initial equilibrium distribution of the 
baths at their respective temperatures and conditioned on the system starting in pure state $\ket{i}$
and finishing in pure state $\ket{f}$.
Derivatives of $G_{if}(\nu)$ with respect to $\nu$ generate moments of the energy change. Here we are
interested in the first derivative
\begin{eqnarray}
\label{eq:average-energy-change}
\left<\Delta E_C\right> &=& \frac{d}{d(i\nu)}G_{if}(\nu)|_{\nu=0}
\end{eqnarray}
Furthermore we are only interested in thermal power, the limit $\frac{1}{t}\left<\Delta E_C\right>$ 
when $t$, the duration of the process, is long. 

Stepping first back a bit, the calculation 
of $G_{if}(\nu)$ proceeds by representing 
$U$ and $U^{\dagger}$ as path integrals. Path integrals
for spins are known in general~\cite{atland-simons},
and have recently been used by one of us to estimate the 
errors in quantum computing~\cite{Aurell2018JSP}.
For the problem at hand a much simpler representation is
however sufficient, where the spin paths $X$ and $Y$ representing 
$U$ and $U^{\dagger}$ are piece-wise constant, taking values $\pm\frac{1}{2}$~\cite{Leggett87}.
The baths are composed of sets of harmonic oscillators
interacting linearly with the spin, and their terms
in $U$ and $U^{\dagger}$ as well as 
$\rho_H^{eq}$, $\rho_C^{eq}$ and $e^{\pm i\nu H_{C}}$
can be represented as standard path integrals,
which can be integrated out as many Gaussians~\cite{FeynmanVernon}.
The functional $G_{if}(\nu)$ can hence be represented as
as a double path integral of the spin paths $X$ and $Y$
weighted by an action, \textit{i.e.} as $e^{\frac{i}{\hbar}{\cal A}[X,Y]}$.
At $\nu=0$ this is the same spin-boson
path integral derived in~\cite{Leggett87}, which
represents the quantum operation of moving the density matrix
of the spin at time zero to the density matrix of the spin at time $t$.
For non-zero values of $\nu$ additional terms appear in ${\cal A}$,
details are summarized in Appendix~\ref{sec:heat-and-NIBA}.

In practice the spin-boson path integrals are quite cumbersome to do without
replying on the ``non-interacting blip approximation'' (NIBA).
The terms in ${\cal A}$ that arise from
integrating out the bath(s) are double integrals with kernels,
and NIBA means that those kernels should have short enough memory.
More precisely, memory should be shorter than the duration of the
periods when $X$ and $Y$ take the same value, $(\frac{1}{2},\frac{1}{2})$ or
$(-\frac{1}{2},-\frac{1}{2})$, so that the bath can only remember the
preceding such period.
Since the switching rate of paths in the double path integral
is given by the 
tunneling rate in the system Hamiltonian, NIBA
is hence expected to hold when that tunneling rate is small.
The same reasoning essentially holds for non-zero values of $\nu$.
The set-up is summarized in Appendices~\ref{app:spin-boson}
and~\ref{sec:heat-and-NIBA}.

With caveats discussed in Appendix~\ref{sec:long-term-limit}
the stationary state (for the spin) in the spin-boson problem
can then (within NIBA) be determined by almost classical arguments.
A transition from the up state $(\frac{1}{2},\frac{1}{2})$ 
to the down state $(-\frac{1}{2},-\frac{1}{2})$ proceeds
through two channels labeled by which spin path goes
first ($X$ or $Y$), and the time ($\Delta t$) spent in the intermediate 
``blip'' state ($(\frac{1}{2},-\frac{1}{2})$ or $(-\frac{1}{2},\frac{1}{2})$).
The first jump occurs with intrinsic rate  
$i\frac{\Delta}{2}$
or $-i\frac{\Delta}{2}$ 
and the second jump with the other rate.
Altogether, for both
kinds of channels, this gives $\frac{\Delta^2}{4}$.
The two baths are in equilibrium with respect to the spin
before the jump, and integrating them out thus 
leads to characteristic functions $S_C$ and $X_C$ for the cold bath
and $S_H$ and $X_H$ for the hot bath.
Summing contributions from all channels thus gives an
overall transition rate from up to down:
\begin{equation}
A = \frac{\Delta^2}{2} \int e^{-\frac{1}{\hbar}(S_C+S_H)} \cos\frac{1}{\hbar}(X_C+X_H-\epsilon\Delta t) \, d\Delta t
\end{equation}
and a similar overall transition rate from down to up
\begin{equation}
D = \frac{\Delta^2}{2} \int e^{-\frac{1}{\hbar}(S_C+S_H)} \cos\frac{1}{\hbar}(X_C+X_H+\epsilon\Delta t)\, d\Delta t
\end{equation}
The stationary probability to be up is $\frac{D}{A+D}$.
This expression is formally identical with the dimensional
arguments in Section~\ref{sec:arguments}:
$A$ may be identified with $\Gamma_{\uparrow\downarrow}$;
and $D$ with $\Gamma_{\downarrow\uparrow}$
\footnote{The sum $A+D$ is proportional to the quantity called $\tilde{g}$
in~\cite{Leggett87} (at zero Laplace transform parameter),
and the difference $D-A$ is proportional to $-\tilde{h}$.
The magnetization is $\frac{D-A}{A+D}$, which equals
$-\tilde{h}/\tilde{g}$ in the notation of~\cite{Leggett87}.
}.

The calculations of thermal power detailed in 
Appendices~\ref{sec:singluar-heat-NIBA}-\ref{sec:long-term-limit-derivative}
rely crucially on exact relations between 
the derivative of the action ${\cal A}$ with respect to
the parameter $\nu$ at $\nu=0$, and the derivatives of 
the two functions $S$ and $X$ with respect to the time argument.
It is then convenient to introduce additional characteristic 
functions of the hot and the cold baths~\footnote{Equivalent 
functions have been introduced in the previous literature, but not 
exactly for these quantities}
\begin{eqnarray}
C_+^C(t) &=& e^{-\frac{1}{\hbar}S_C + \frac{i}{\hbar}X_C} \\ 
C_+^H(t) &=& e^{-\frac{1}{\hbar}S_H + \frac{i}{\hbar}X_H} \\ 
C_-^C(t) &=& e^{-\frac{1}{\hbar}S_C - \frac{i}{\hbar}X_C} \\ 
C_-^H(t) &=& e^{-\frac{1}{\hbar}S_H - \frac{i}{\hbar}X_H}  
\end{eqnarray}
The quantity $A$ introduced above is then 
\begin{equation}
A = \frac{\Delta^2}{4}\int \left(C_+^C(t)C_+^H(t)
e^{-i\frac{\epsilon t}{\hbar}}+C_-^C(t)C_-^H(t)e^{i\frac{\epsilon t}{\hbar}}\right)\, dt
\end{equation}
and similarly for $D$.

As determined in appendix,
the rate of energy change in the cold bath
while the system is respectively in the up and the down state can be written,
compare  \eqref{eq:a-result-two-baths},                     
\begin{eqnarray}
\label{eq:power-in-up}
\pi_{up} &=& -i\hbar \frac{\Delta^2}{4} \int dt e^{-i\frac{\epsilon t}{\hbar}}\frac{dC_+^C(t)}{dt} C_+^H(t) \nonumber \\  
                   &&    +i\hbar \frac{\Delta^2}{4} \int dt e^{i\frac{\epsilon t}{\hbar}}\frac{dC_-^C(t)}{dt} C_-^H(t) \\
\label{eq:power-in-down}
\pi_{down} &=& -i\hbar \frac{\Delta^2}{4} \int dt e^{\frac{i\epsilon t}{\hbar}}\frac{dC_+^C(t)}{dt} C_+^H(t) \nonumber \\ 
                     &&  +i\hbar\frac{\Delta^2}{4}  \int dt e^{-\frac{i\epsilon t}{\hbar}}\frac{dC_-^C(t)}{dt} C_-^H(t)
\end{eqnarray}
An interpretation of the above results is
that $C_+^C$, $C_+^H$, $C_-^C$ and $C_-^H$
are the influence functionals from integrating out
the baths when the forward and backward paths
of the spin are fixed and opposite.
These influence functionals are of the form
$\hbox{Tr}\left[U\rho^{eq,\uparrow}V^{\dagger}\right]$
with different unitary operators 
applied to the left and to the right. 
Differentiating $U$ and $V^{\dagger}$
with respect to
time brings down $-\frac{i}{\hbar}\left(H_B+H_I\right)$
and
 $\frac{i}{\hbar}\left(H_B+H_I'\right)$
with different interaction Hamiltonians
on the two sides because the spin coordinate is different on the two sides.
The bath Hamiltonians are however the same
and their contributions hence cancel, and 
the remaining terms are 
expectation values of the interaction Hamltonians,
conditional on which state the spin started from, which 
path jumped first, and the blip duration.
In this way \eqref{eq:power-in-up} and
\eqref{eq:power-in-down}
can be seen to give an estimate of the type of
\eqref{eq:power-strong-coupling}.

\section{The polaron transform picture}
\label{sec:polaron-transform}
Another interpretation of the
results in \eqref{eq:power-in-up} and \eqref{eq:power-in-down}
is based on the polaron transform.
Changing $\hat{\sigma}_z$ from up to down has the same effect on 
the bath energy as instantaneously shifting the position of every
bath oscillator $q_b$ by an amount $2\frac{C_b}{m_b\omega_b^2}$. 
Such a shift is generated by 
$\hat{B}_{+}=\exp\left(i 2\sum_b\frac{C_b}{\hbar m_b\omega_b^2}\hat{p}_b\right)$
where $\hat{p}_b$ is the momentum operator of oscillator $b$.
Similarly $\hat{B}_-=\exp\left(-i 2\sum_b\frac{C_b}{\hbar m_b\omega_b^2}\hat{p}_b\right)$
has the same effect on the bath energy as changing $\hat{\sigma}_z$ from down to up.

The function $C_-(t)=e^{-\frac{1}{\hbar}S - \frac{i}{\hbar}X}$ for the cold or hot bath ($C$ or $H$)
is therefore the same as $\left<\hat{B}_-(t)\hat{B}_+(0)\right>_{eq}$
where the operators are in Heisenberg picture, and the average is over the bath in equilibrium.
Similarly $C_+(t)=e^{-\frac{1}{\hbar}S + \frac{i}{\hbar}X}$ is the same as $\left<\hat{B}_-(0)\hat{B}_+(t)\right>_{eq}$.
The effective jump rates are thus 
\begin{eqnarray}
A &=& \frac{\Delta^2}{4}\int \left<\hat{B}_-(0)\hat{B}_+(t)\right>_{C,eq} \left<\hat{B}_-(0)\hat{B}_+(t)\right>_{H,eq} e^{-i\frac{\epsilon t}{\hbar}}\nonumber \\
&& + \left<\hat{B}_-(t)\hat{B}_+(0)\right>_{C,eq} \left<\hat{B}_-(t)\hat{B}_+(0)\right>_{H,eq}
e^{i\frac{\epsilon t}{\hbar}} \, dt
\end{eqnarray}
and similarly for $D$.
The above may be used to derive the weak-interaction limit, since then
$\hat{B}_+\approx \mathbf{1} + i 2\sum_b\frac{C_b}{\hbar m_b\omega_b^2}\hat{p}_b$
and $\hat{B}_-\approx \mathbf{1} - i 2\sum_b\frac{C_b}{\hbar m_b\omega_b^2}\hat{p}_b$,
and (linear terms cancel)
\begin{equation}
\left<\hat{B}_-(t)\hat{B}_+(0)\right>_{eq} \approx 1 + \frac{4}{\hbar^2}\sum_b \frac{C_b^2}{m_b^2\omega_b^4}\left<\hat{p}_b(0)\hat{p}_b(t)\right>_{eq}
\end{equation}
Except for $\epsilon$ very small this gives
the effective jump rate proportional to the
sum of the spectral powers of the cold and hot bath at frequency $\epsilon/\hbar$,
which can be compared \textit{e.g.} to~\cite{KarimiPekola2016} (Eq.~3).

In a similar manner one may also consider \eqref{eq:power-in-up} and \eqref{eq:power-in-down}. 
The derivatives $\frac{dC_+^C(t)}{dt}$ and $\frac{dC_-^C(t)}{dt}$ translate (in weak coupling) to
$\frac{4}{\hbar^2}\sum_b \frac{C_b^2}{m_b^2\omega_b^4}\left<\hat{p}_b(0)\frac{d\hat{p}_b(t)}{dt}\right>_{C,eq}$ and
$\frac{4}{\hbar^2}\sum_b \frac{C_b^2}{m_b^2\omega_b^4} \left<\frac{d\hat{p}_b(t)}{dt}\hat{p}_b(0)\right>_{C,eq}$.
The dependence on the hot bath is only to higher orders in the interaction coefficients, and therefore drops out.
Given that $C_+^C(0)$ and $C_-^C(0)$ are both equal to one, one may integrate by parts,
which gives
\begin{eqnarray}
\label{eq:power-in-up-weak}
\pi_{up} &\approx&  \frac{\Delta^2}{4} \epsilon\int dt e^{-i\frac{\epsilon t}{\hbar}} C_+^C(t)   \nonumber \\  
                   && \quad+ \frac{\Delta^2}{4}\epsilon \int dt e^{i\frac{\epsilon t}{\hbar}} C_-^C(t) \\
\label{eq:power-in-down-weak}
\pi_{down} &\approx& -\frac{\Delta^2}{4} \epsilon \int dt e^{\frac{i\epsilon t}{\hbar}} C_+^C(t) \nonumber \\ 
                     &&\quad -  \frac{\Delta^2}{4}   \int dt e^{-\frac{i\epsilon t}{\hbar}} C_-^C(t)
\end{eqnarray}
which is of the same form as
\eqref{eq:power-weak-coupling-up}
and
\eqref{eq:power-weak-coupling-down}.

\section{Discussion}
\label{sec:discussion}
In this paper we have considered thermal power (heat per unit time) 
through a qubit interacting with two or several baths
as in the spin-boson problem~\cite{Leggett87}. 
By an 
extension of the Feynman-Vernon influence 
functional method it is possible to
compute the 
distribution of energy changes in a bath or baths of
harmonic oscillators interacting with a general quantum 
system~\cite{AurellEichhorn,Aurell2018b,FunoQuan2018,Carrega2016}.
Here we have adapted this approach to the situation where the system in one spin.

The advantage of the Feynman-Vernon method is that while each oscillator
in the bath is only perturbed slightly, and the system-bath interaction 
hence assumed linear in the harmonic oscillator coordinates,
the accumulated effect on the system from all the bath oscillators
can be large. A Feynman-Vernon theory of energy changes in a bath 
is thus a way to model quantum heat in a system interacting strongly with 
its environment. In this paper we have only considered the expected 
value, but in principle higher moments can also be computed \textit{e.g.}
by the formulae given in~\cite{Aurell2018b}.
Furthermore we have only considered the stationary case (constant drive)
and the long-time limit which can be analyzed by Laplace transforms,
as was already done in~\cite{Leggett87}. 

If an assumption analogous to the ``non-interacting blip approximation'' (NIBA)
is made, the general structure of the answer is quite simple, and basically
follows by dimensional arguments. It can also be expressed in terms of 
correlation functions and time derivatives of correlation functions
after a polaron transform.
While the final result is simple, the intermediate calculations are not,
as seems to be the case for most path integral treatments of the spin-boson problem,
compare~\cite{Leggett87} as well as the later 
literature~\cite{Weiss-book,GrifoniHanggi1998,Grifoni1997,HartmannGoychukGrifoniHanggi2000}.
For the quantum state a much simpler approach is possible
using the polaron transform directly~\cite{Aslangul1986,Dekker1987}. Since our 
result for thermal power can
also be expressed in terms of quantities after a polaron transform, it would
be interesting to know if it can also be found in a simpler manner. We
leave this question to future work,
as well as numerical determination terms~\eqref{eq:power-in-up} 
and~\eqref{eq:power-in-down} in thermal power.

We end by noting that for a qubit interacting with two baths 
the prediction of NIBA may be not only incorrect, but also
physically inadmissable. The limits of validity of NIBA
may thus be qualitatively different in non-equilibrium compared to equilibrium.
This question deserves further study.
We further note that in NIBA the condition 
that thermal power to the cold bath be positive 
appears different than the admissibility condition on the state. 
Conceivably there may hence be situations where NIBA is appropriate,
for the quantum state but not for quantum thermodynamics.
This issue also deserves further study.

\section*{Acknowledgments}
\label{sec:ack}
This work was supported by ESPCI Chaire Joliot 2018 (EA).
EA thanks Jukka Pekola and Bayan Karimi for many
discussions on heat flows in superconducting
devices, Yuri Galperin for a critical reading of
the MS, and Dmitry Golubev for showing results
prior to publication. Results equivalent 
to Eq.~\eqref{eq:power-in-up} and~\eqref{eq:power-in-down}
have also been derived independently by Golubev in the case of zero bias. 
FM was supported by an Erasmus+ Student Mobility for
Traineeship (Politecnico di Torino, Italy),
and thanks Nordita (Stockholm, Sweden) for hospitality.

\bibliography{fluctuations,quantum-fluctuations}

\appendix

\begin{widetext}
\section{Summary of spin-boson theory and NIBA}
\label{app:spin-boson}
The calculations in Section~\ref{sec:strong-coupling} are for the quantum thermal power
and two baths
what Leggett and collaborators did in the 80ies for the development of the quantum state and one bath~\cite{Leggett87}.
This Appendix summarizes relevant results from that earlier calculation.
For ease of comparison (here and in later related Appendices)
we follow the notation of~\cite{Leggett87}. 
We restate the system (qubit) Hamiltonian: 
\begin{equation}
H_S = -\hbar\frac{\Delta}{2} \hat{\sigma}_x +  \frac{\epsilon}{2} \hat{\sigma}_z 
\label{eq:H_S}
\end{equation}
where $\Delta$ is a rate (dimension $\left(\hbox{time}\right)^{-1}$), and $\epsilon$ is the level splitting.
The bath Hamiltonians are, in classical notation,
\begin{eqnarray}
\label{eq:H_C}
H_C &=& \sum_{b\in C} \frac{p_b^2}{2m_b}+\frac{1}{2}m_b\omega_b^2 q_b^2 \\
\label{eq:H_H}
H_H &=& \sum_{b\in H} \frac{p_b^2}{2m_b}+\frac{1}{2}m_b\omega_b^2 q_b^2 
\end{eqnarray}
where the parameters $m_b$ and $\omega_b$ are the mass and angular frequency of each oscillator
and $C$ and $H$ also stand for the sets of oscillators in respectively the cold bath and the hot bath.
The system-bath interactions are similarly 
\begin{eqnarray}
\label{eq:H_LS}
H_{CS} &=& -\sum_{b\in L} C_b q_b \hat{\sigma}_z \\
\label{eq:H_RS}
H_{HS} &=& -\sum_{b\in R} C_b q_b \hat{\sigma}_z 
\end{eqnarray}
where $C_b$ is the interaction coefficient between bath oscillator $b$
and the qubit, and $\hat{\sigma}_z$ operates on the qubit.
The coupling coefficients $C_b$ have 
dimension $(\hbox{energy})\cdot (\hbox{length})^{-1}$.

The Feynman-Vernon transition probability of a general quantum system interacting with two baths is
\begin{equation}
\label{P-if}
P_{if} = \hbox{Tr}_{CH}
\matrixel{f}{U \left(\dyad{i} \oplus \rho_{CH}^{eq}\right) U^{\dagger}  }{ f} 
\end{equation}
where the initial state of the baths $\rho_{CH}^{eq}$ is the product state of two thermal states
$\rho_{C}^{eq}$ and $\rho_{H}^{eq}$, at two temperatures.
$U$ is the big unitary expressing the forward time evolution due to
the total Hamiltonian given by
\eqref{eq:H_S},
\eqref{eq:H_C}
\eqref{eq:H_H}
\eqref{eq:H_LS}
and
\eqref{eq:H_RS},
and $U^{\dagger}$ (the adjoint) is the backward time evolution.

The bath coordinates in \eqref{P-if} can be integrated out to yield
\begin{equation}
\label{P-if-v2}
P_{if} = \int_{if} {\cal D}X {\cal D}Y e^{\frac{i}{\hbar}S_S[X]-\frac{i}{\hbar}S_S[Y]+\frac{i}{\hbar}S^{C}_i[X,Y]+\frac{i}{\hbar}S^{H}_i[X,Y]-\frac{1}{\hbar}S^{C}_r[X,Y]-\frac{1}{\hbar}S^{H}_r[X,Y]} 
\end{equation}
where $X$ is the system coordinate in the forward system path (part of the representation of $U$),
$Y$ is the system coordinate in the backward system path (part of the representation of $U^{\dagger}$),
and $\int_{if}$ denotes the projection on initial and final states (integrals over initial and final positions of the system in the
forward and backward path). 
The result of integrating out the cold bath is
$\frac{i}{\hbar}S^{C}_i[X,Y]-\frac{1}{\hbar}S^{C}_r[X,Y]$,
and the result of integrating out the hot bath is
$\frac{i}{\hbar}S^{H}_i[X,Y]-\frac{1}{\hbar}S^{H}_r[X,Y]$.
The real terms ($S_r$) depend on the difference $X-Y$ at two different times
while the imaginary terms ($S_i$) depend on the difference $X-Y$ at a later time,
and the sum $X+Y$ at an earlier time.

For the system and bath interaction described by 
\eqref{eq:H_S},
\eqref{eq:H_C}
\eqref{eq:H_H}
\eqref{eq:H_LS}
and
\eqref{eq:H_RS} the system paths $X$ and $Y$ can be represented as piece-wise constant,
taking value $\frac{1}{2}$ when the spin is up, and $-\frac{1}{2}$ when the spin is down.  
This means that at any one one time the forward-backward system path pair can take only 
four values  $(\frac{1}{2},\frac{1}{2})$, $(-\frac{1}{2},-\frac{1}{2})$,
$(\frac{1}{2},-\frac{1}{2})$ and $(-\frac{1}{2},\frac{1}{2})$.
The two first are in the terminology of~\cite{Leggett87} called \textit{sojourns}
and correspond to populations, the diagonal elements of the density matrix, up and down.
The last first are in the terminology of~\cite{Leggett87} called \textit{blips}
and correspond to coherences, the off-diagonal elements of the density matrix.
The kind of sojourn and blip can be indicated by variables
$\chi=X+Y$ and $\xi=X-Y$, both taking values $\pm 1$. 
A given double path in $X$ and $Y$, conventionally starting from the up sojourn, can therefore be represented as
\begin{equation}
\mathbf{\sigma}= (\chi_0=1,\Delta t_0,\xi_1,\Delta t_1,\chi_1,\Delta t_2,\xi_2,\Delta t_3,\ldots)
\end{equation}
where $\Delta t_0, \Delta t_2,\ldots$ are the durations of the sojourns and $\Delta t_1, \Delta t_3,\ldots$ are the durations of the blips.
The first sojourn starts at time $t_0$ and the $n$'th sojourn at time $t_{2n}=t_0+\sum_{j=0}^{2n-1} \Delta t_j$;
the $n$'th blip starts at time $t_{2n+1}=t_0+\sum_{j=0}^{2n} \Delta t_j$.

The $\hat{\sigma}_x$ terms in \eqref{eq:H_S} translate to weights in the integrations ${\cal D}X$ and ${\cal D}Y$
in \eqref{P-if-v2} which are $+i(\frac{\Delta}{2})$ if the forward path ($X$) jumps, and $-i(\frac{\Delta}{2})$ if the backward path ($Y$) jumps.
Everything else is included in the
total exponent in \eqref{P-if-v2} which one can write as
\begin{eqnarray}
{\cal S}(\mathbf{\sigma}) &=& \sum_j (-i\frac{\epsilon}{\hbar})\Delta t_{2j-1}-\frac{1}{\hbar} \left(S^C_j(\Delta t_{2j-1})+S^H_j(\Delta t_{2j-1})\right)
                              +\frac{i}{\hbar}\xi_j \chi_{j-1} \left(X^C_{j,j-1}(t_{2j-2},t_{2j-1},t_{2j})+X^H_{j,j-1}(t_{2j-2},t_{2j-1},t_{2j})\right) \nonumber \\
&+& \sum_j\sum_{k<j} -\frac{1}{\hbar} \xi_j\xi_k\left(\Lambda^C_{jk}(t_{2j},t_{2j+1},t_{2k},t_{2k+1})+\Lambda^H_{jk}(t_{2j},t_{2j+1},t_{2k},t_{2k+1})\right) \nonumber \\
\label{eq:Leggett-formula}
&+&\sum_j\sum_{k<j-1} \frac{i}{\hbar}\xi_j\chi_k \left(X^C_{jk}(t_{2j},t_{2j+1},t_{2k-1},t_{2k})+X^H_{jk}(t_{2j},t_{2j+1},t_{2k-1},t_{2k})\right)
\end{eqnarray}
where all terms are integrals over time of the terms in the exponent in
\eqref{P-if-v2}.   
The first line in above hence represent the terms
$ \frac{i}{\hbar}S_S[X]-\frac{i}{\hbar}S_S[Y]$ which have only one time integral, and which are non-zero only for blips,
the terms $-\frac{1}{\hbar}S^{C}_r[X,Y]-\frac{1}{\hbar}S^{H}_r[X,Y]$, with both terms
in the same blip, and $\frac{i}{\hbar}S^{C}_i[X,Y]+\frac{i}{\hbar}S^{H}_i[X,Y]$
with the sojourn immediately before the blip.
The second and third line in \eqref{eq:Leggett-formula} correspond to times separated by at least one sojourn.

The \textit{Non-interacting blip approximation} (NIBA) of~\cite{Leggett87} is to
ignore the second and third line of \eqref{eq:Leggett-formula}, and to assume that 
$X^C_{j,j-1}$ and $X^H_{j,j-1}$ only depend on the associated blip duration $\Delta t_{2j-1}$.
The validity of this approximation was discussed in depth in~\cite{Leggett87} and
in the later literature, see \textit{e.g}~\cite{Weiss-book,GrifoniHanggi1998,Grifoni1997,HartmannGoychukGrifoniHanggi2000}.
Here we only note that it is essentially an expansion in small
tunneling rates $\Delta$, as lucidly explained in~\cite{Aslangul1986} and~\cite{Dekker1987},
with long blip durations suppressed as a result of the interaction between the system and the baths.

The content of NIBA is thus expressed in the following two characteristic functions of the baths,
which we write for the cold bath as 
\begin{eqnarray}
\label{eq:X-C}
X^C_{j,j-1}(\Delta t_{2j-1})      &=&  \sum_{b\in C} \frac{C_b^2}{2m_b\omega_b^3} \sin\omega_b\Delta t_{2j-1} \\ 
\label{eq:S^C}
S^C_j(\Delta t_{2j-1})  &=& \sum_{b\in C} \frac{C_b^2}{2m_b\omega_b^3} \coth\left(\frac{\omega_b\hbar\beta_C}{2}\right)(1-\cos\omega_b \Delta t_{2j-1}) 
\end{eqnarray}
In above the sums are over oscillators in the cold bath and
$\beta_C$ is the inverse temperature of the cold bath. The formulas for the contributions from the hot bath are analogous.

It is customary to also write the above functions as $Q_1$ and $Q_2$
as these are equivalent in NIBA. If one does not assume NIBA, $X_{j,j-1}$ would however be the sum of three terms
$Q_1$ with different arguments, where the one above is the shortest time.

\section{Heat and NIBA}
\label{sec:heat-and-NIBA}
The starting point is the generating function of energy changes in the cold bath
\begin{eqnarray}
\label{eq:FCS}
G_{if}(\nu) &=& \hbox{Tr}_{CH}\matrixel{f}{e^{i\nu H_{C}} U \left(\dyad{i} \oplus \rho_{CH}^{eq} e^{-i\nu H_{C}} \right)  U^{\dagger} }{ f}
\end{eqnarray}
This equation is the same as \eqref{P-if} above, except that 
exponentials of the Hamiltonian of the cold bath have been inserted at
the initial and final time.
It is assumed in \eqref{eq:FCS} that  $e^{-i\nu H_{C}}$ commutes with the
initial density matrix of the baths $\rho_{CH}^{eq}$; this issue, related
to strong coupling, will be discussed below.

As for \eqref{P-if} we can introduce path integral representations of $U$ and $U^{\dagger}$
and integrate out the two baths.
The result must analogously to \eqref{P-if-v2} look like
\begin{equation}
\label{G-if}
G_{if}(\nu) = \int_{if} {\cal D}X {\cal D}Y e^{\frac{i}{\hbar}S_S[X]-\frac{i}{\hbar}S_S[Y]+\frac{i}{\hbar}\left(S^C_i[X,Y]+S^H_i[X,Y]\right)-
\frac{1}{\hbar}\left(S^C_r[X,Y]+S^H_r[X,Y]\right) +\frac{i}{\hbar} {\cal J}_{\nu}[X,Y] + \frac{i}{\hbar} \tilde{\cal J}_{\nu}[X,Y]}
\end{equation}
where the two new functionals ${\cal J}_\nu$ and $\tilde{\cal J}_\nu$,
which represent the distribution of energy changes in the cold bath, are quadratic in $X$ and $Y$.
The two terms are for later convenience separated as to
and respectively depending anti-symmetric and symmetric combinations in the exchange of times.
In earlier contributions the same two functionals and their kernels
were distinguished by superscripts $(2)$ and $(3)$~\cite{AurellEichhorn,Aurell2017,Aurell2018a}.
Here we choose to streamline the formalism, additionally because the similar functional with superscript $(1)$
does not appear; for a discussion, see~\cite{Aurell2017}.

When $\nu$ is equal to zero $G_{if}(\nu)$ is equal to $P_{if}$, and the two  functionals ${\cal J}_\nu$ and $\tilde{\cal J}_\nu$ must vanish.
In this paper we are concerned with the terms linear in $\nu$ which are given by
\begin{eqnarray}
\label{J-linear-order}
{\cal J}_{\nu}[X,Y]       &\approx & \nu \int^{t_f}_{t_i}dt \int^{t}_{t_i}ds \left(X_tY_s - X_s Y_t\right)h(t-s) \\
\label{tildeJ-linear-order}
\tilde{\cal J}_{\nu}[X,Y] &\approx & \nu \int^{t_f}_{t_i}dt \int^{t}_{t_i}ds \left(X_tY_s + X_s Y_t\right)\tilde{h}(t-s) 
\end{eqnarray}
with two kernels 
\begin{eqnarray}
\label{eq:h2-definition}
h (t-s)&=& i\hbar\sum_{b\in C }\frac{C_b^2}{2m_b}\coth(\frac{\beta\hbar\omega_b}{2})\sin\omega_b(t-s) \\
\label{eq:h3-definition}
\tilde{h}(t-s) &=& \hbar\sum_{b\in C}\frac{C_b^2}{2m_b}\cos\omega_b(t-s)
\end{eqnarray}
These two kernels are the same as $h^{(2)}$
and $h^{(3)}$ in \cite{AurellEichhorn}, except
for a factor $\hbar$.

It is a non-trivial fact~\cite{AurellEichhorn} that 
$h$ and $\tilde{h}$ are proportional to 
time derivatives of the Feynman-Vernon kernels 
\begin{eqnarray}
\label{eq:tildeh-k_i}
\tilde{h}(\tau) &=& \hbar\frac{d}{d\tau}k_i(\tau) \qquad k_i = \sum_b \frac{C_b^2}{2m_b\omega_b} \sin\omega_b\tau \\
\label{eq:h-k_r}
h(\tau) &=& -i\hbar\frac{d}{d\tau}k_r(\tau) \qquad k_r = \sum_b \frac{C_b^2}{2m_b\omega_b} \coth\left(\frac{\omega\hbar\beta}{2}\right)\cos\omega_b\tau  
\end{eqnarray}
Similar relations between second integrals of these kernels will be crucial in the following.

We can now represent 
$G_{if}(\nu)$ in a similar way to \eqref{eq:Leggett-formula}
with new terms stemming from 
$J$ and $\tilde{J}$.
We can write these as
\begin{eqnarray}
\label{eq:J-Leggett-way}
J(\mathbf{\sigma}) &\approx& \frac{1}{2}\nu\sum_{j}\sum_{k<j} \xi_j\chi_k X^{(1)}_{jk} - \nu\frac{1}{2}\sum_{j}\sum_{k \geq j} \xi_j\chi_k X^{(1)}_{jk} \\
\label{eq:tildeJ-Leggett-way}
\tilde{J}(\mathbf{\sigma}) &\approx& -\frac{1}{2}\nu\sum_{j} S^{(1)}_j +\frac{1}{2}\nu\sum_{j} S^{(1')}_j \nonumber \\
&& -\frac{1}{2}\nu\sum_{j}\sum_{k<j} \xi_j\xi_k \Lambda^{(1)}_{jk} + \frac{1}{2}\nu\sum_{j}\sum_{k<j} \chi_j\chi_k \Lambda^{(1')}_{jk} 
\end{eqnarray}
In above $X^{(1)}_{jk}$ are the first-order terms in $\nu$ from the kernels anti-symmetric in the time exchange.
In contract to the imaginary Feynman-Vernon kernel, both the blip-sojourn and sojourn-blip terms appear.
Furthermore  $S^{(1)}_j$ and $S^{(1')}_j$ are the first-order terms in $\nu$ from the kernels symmetric
in the time exchange where both times fall in the same time interval.
In contrast to the real Feynman-Vernon kernel, there are such terms from both blips and sojourns.
Finally $\Lambda^{(1)}_{jk}$ and $\Lambda^{(1')}_{jk}$ are terms from two intervals
of the same kind, either two blips or two sojourns.

A NIBA-like approximation to \eqref{eq:J-Leggett-way} means to include only 
the terms from an adjacent blip and sojourn. 
These are on the one hand terms like 
$-\frac{1}{2}\nu \xi_j\chi_{j-1} X^{(1)}_{j,j-1}$, and on the other
$\frac{1}{2}\nu \xi_j\chi_{j} X^{(1)}_{j,j}$
both of which depend on time increments as discussed 
for $X_{j,j-1}$ above. Only one of these time increments is
for a blip interval (the same blip interval), and we are therefore led to
\begin{equation}
X^{(1)}_{j,j-1}\approx X^{(1)}_{j,j}\approx K(\Delta t_{2j-1}) \equiv i\hbar\sum_{b\in C }\frac{C_b^2}{2m_b\omega_b^2}\coth(\frac{\beta\hbar\omega_b}{2})\sin\omega_b\Delta t_{2j-1}
\end{equation} 
From this we have the NIBA-like approximation
\begin{eqnarray}
\label{eq:J-NIBA}
J(\mathbf{\sigma}) &\approx& \frac{1}{2}\nu\sum_{j}\xi_j(\chi_{j-1}-\chi_j) K(\Delta t_{2j-1})
\end{eqnarray}
Comparing to \eqref{eq:h-k_r} and \eqref{eq:S^C} we see that
\begin{equation}
\label{eq:K-and-S}
K(\tau) = i\hbar\frac{d}{d\tau} S^C_j(\tau)  
\end{equation}

A NIBA-like approximation to \eqref{eq:tildeJ-Leggett-way}
is a bit more involved, for two reasons.
First the two terms $S^{(1)}_j$ and $S^{(1')}_j$ both need to be included, and they are both diverging in the bath cut-off frequency.
This requires a separate discussion which we give below in Appendix~\ref{sec:singluar-heat-NIBA}.
Second, the terms on the second line of \eqref{eq:tildeJ-Leggett-way}
cannot be neglected entirely.
This is so because the interaction of two neighboring
sojourns ($\frac{1}{2}\nu \chi_j\chi_{j-1} \Lambda^{(1')}_{j,j-1}$)
has one terms which depends on the intervening blip time,
and which hence gives
\begin{equation}
\Lambda^{(1')}_{j,j-1} \approx \tilde{K}(\tau) \equiv -\hbar\sum_{b\in C }\frac{C_b^2}{2m_b\omega_b^2}\cos\omega_b\tau
\end{equation} 
Comparing to \eqref{eq:X-C} we see that
\begin{equation}
\label{eq:tildeK-and-X}
\tilde{K}(\tau) = -\hbar\frac{d}{d\tau} X^C_{j,j-1}(\tau) 
\end{equation}

\section{Ohmic baths}
\label{app:Ohmic-baths}
Ohmic baths have spectra (density of states)
that are continuous up to some very large upper cut-off $\Omega$
and increase quadratically with frequency. The number of oscillators with frequencies in the
interval $[\omega,\omega+d\omega]$ is $f(\omega)d\omega$ can then be taken to be
\begin{equation}
\label{eq:f-sharp}
 \begin{array}{lcll}
     f(\omega) &=& \frac{2}{\pi}\omega_c^{-3}\omega^2 \qquad &\omega<\Omega \\
     f(\omega) &=& 0                            & \omega>\Omega 
 \end{array}    
\end{equation}
where $\omega_c$ is some characteristic frequency less than $\Omega$.
The total number of oscillators is then $\frac{2}{3\pi}\left(\frac{\Omega}{\omega_c}\right)^3$.

An alternative version is to take a smooth cut-off:
\begin{equation}
\label{eq:f-exp}
     f(\omega) = \frac{2}{\pi}\omega_c^{-3}\omega^2 \exp\left(-\frac{\omega}{\Omega}\right)
\end{equation}
In this case the number of bath oscillators is  
$\frac{12}{\pi}\left(\frac{\Omega}{\omega_c}\right)^3$.

The system-bath interactions are characterized by 
two parameters $\eta_C$ and $\eta_H$ 
such that for an oscillator in the cold bath
\begin{equation}
\label{eq:C-cL}
C_{b}=\sqrt{\omega_c^{3}m_{\omega}\eta_L}
\end{equation}
and for an oscillator in the hot bath
\begin{equation}
\label{eq:C-cR}
C_{b}=\sqrt{\omega_c^{3}m_{\omega}\eta_R}
\end{equation}
For the spin-coupling problem
the dimensions of 
$\eta_C$ and $\eta_H$ are 
$(\hbox{mass})\cdot (\hbox{length})^{2}\cdot (\hbox{time})^{-1}$
\textit{i.e.} the action.

The terms $X_{j,j-1}(\tau)$ and $S_j(\tau)$ in
\eqref{eq:X-C} and \eqref{eq:S^C} were computed in~\cite{Leggett87}
as $\eta \tan^{-1} (\Omega\tau)$,
and
$\frac{1}{2}\eta \log(1+\Omega^2\tau^2) + \eta \log\left(\frac{\hbar\beta}{\pi\tau}\sinh\frac{\pi\tau}{\hbar\beta}\right)$.
The first is essentially a sign function.
The second starts as
$\frac{\eta}{2}\Omega^2\tau^2$ in the interval $\tau<<\Omega^{-1}$,
then grows as 
$\eta \log \Omega\tau + \frac{\eta}{2\Omega^2\tau^2}$
in the interval 
$\Omega^{-1}<<\tau<<\hbar\beta$
and finally 
behaves as 
$\eta \log\frac{\Omega\hbar\beta}{2\pi} +\frac{\eta\pi}{\hbar\beta}|\tau|$
when $\tau>>\hbar\beta$.
The derivative  $\partial_{\beta} S_j(\tau)$
evaluates to $\eta/\beta (1- \frac{\pi\tau}{\hbar\beta}\coth\frac{\pi\tau}{\hbar\beta})$.
which is always negative.  $S_j(\tau)$ is hence an increasing
function of bath temperature.
The second derivative  $\partial_{\beta\tau} S_j(\tau)$
evaluates to $ \frac{\pi\eta}{\hbar\beta^2}(-\coth\frac{\pi\tau}{\hbar\beta}
+\frac{\pi\tau}{\hbar\beta}\sinh^{-2}\frac{\pi\tau}{\hbar\beta})$.
which is also always negative.  $\partial_t S_j(\tau)$ is hence 
also an increasing
function of bath temperature.

$K$ and $\tilde{K}$ can be computed from  
\eqref{eq:K-and-S}
and 
\eqref{eq:tildeK-and-X}:
$\tilde{K}$ is essentially a delta function
on the bath cut-off frequency scale $\Omega^{-1}$,
while
$K$
is basically a delta function on the time scale $\hbar\beta$, and for large $\tau$
a constant.

\section{Interaction through bath momentum}
\label{sec:bath-momentum-coupling}
\begin{theorem}
Let a system described by coordinate $X$ interact with 
by a bath of harmonic oscillators described by coordinate and momenta
$(q_b,p_b)$ through a combined bath and interaction Hamiltonian
$\sum_b \frac{1}{2m_b}\left(p_b + m_b C_b X\right)^2 + \frac{1}{2}m_b\omega_bq_q^2$.
The coupling coefficients $C_b$ vanish at the beginning and the end of the process.
Then the generating function of the change of bath
energy is the same is if the combined bath and interaction 
Hamiltonian would have been $\sum_b \frac{1}{2m_b}p_b^2 + \frac{1}{2}m_b\omega_b^2\left(q_q-\frac{C_b}{\omega_b} X\right)^2$. 
\end{theorem}

The proof proceeds by adapting the calculation in~\cite{Aurell2018b}, in the following steps.

\begin{enumerate}

\item The action corresponding to the Hamiltonian coupled through momentum is
$\int \frac{1}{2}m_b \dot{q}^2_b- m_b C_b X  \dot{q}_b - \frac{1}{2}m_b\omega_bq_q^2$.
By an integration by parts the term linear in  $\dot{q}_b$ is changed to
$\hbox{\textit{boundary terms}} + \int m_b \frac{d}{dt}(C_b X)  q_b$.

\item The path integral of the bath oscillator with fixed initial and final positions can then be considered
to be that of a Lagrangian $\int \frac{1}{2}m_b \dot{q}^2_b 
- \frac{1}{2}m_b\omega_b^2 q_b^2 + m_b\frac{d}{dt}(C_b X)q_b$.
This path integral can then be done as in Feynman-Vernon theory
giving integrals of the external drive (here $ m_b\frac{d}{dt}(C_b X)$) multiplying the 
initial and final positions of the oscillator, and a constant.

\item The integrals are of the type ($u$ in the notation 
of~\cite{Aurell2018b}, Appendix~A) $\frac{1}{\sin\omega_b t}\int^t_0 \sin\omega_b (t-s)  m_b\frac{d}{ds}(C_b X)(s) ds$.
By a partial integration they can be combined with the boundary terms to give
$\frac{m_b\omega_b}{\sin\omega_b t}\int^t_0 \cos\omega_b (t-s)  C_b X ds$, multiplying
the initial position of the bath oscillator in the forward path.
There are four terms of this type with two sign changes compared to \cite{Aurell2018b}, Appendix~A.

\item The constant ($B$ in the notation of~\cite{Aurell2018b}, Appendix~A)
is two terms of the type $\frac{1}{m_b\omega_b\sin\omega_b t}\int^t_0\int^s \sin\omega_b (t-s)\sin\omega_b s'  
m_b\frac{d}{ds}(C_b X)(s) m_b\frac{d}{ds'}(C_b X)(s')   ds'ds$.
By two integrals by parts the sines are turned into cosines multiplying
$(C_b X)(s) (C_b X)(s')$, and there is a change of sign. 
Additionally there is a boundary term $-\frac{m_b}{2}\int C_b^2 X^2 ds$,
the same as appears in the complete square
$-\frac{1}{2}m_b\omega_b^2\left(q_q-\frac{C_b}{\omega_b} X\right)^2$.

\item The integration over the initial and final coordinates of the bath oscillator
proceeds as in~\cite{Aurell2018b}, Appendix~A, and gives in fact the same
result, with $m_bC_b\omega_b$ appearing instead of $C_b$.
One of the authors (E.A.) points out that there is an error in 
Eqs (25) and (A14) in~\cite{Aurell2018b}: the constant appearing in
the kernel ${\cal J}^{(2)}$ should read $(yz'-y'z)/\Delta$ (instead of $(y'z'-yz)/\Delta$).
To linear order in the parameter $\nu$ these two quantities are however the same,
hence there is no difference to the present paper.

 \end{enumerate}
In summary, the only difference to coupling through coordinate is hence that if the coupling coefficient to bath momentum 
is $C$, then the equivalent coupling coefficient to bath coordinate is $m \omega C$,
as is also required dimensionally.

\section{The singular NIBA heat terms}
\label{sec:singluar-heat-NIBA}
In this appendix we estimate the contributions
$S^{(1)}_j$ and $S^{(1')}_j$ 
to \eqref{eq:tildeJ-Leggett-way}.
Both these terms are second integrals of the kernel $\tilde{h}$ in \eqref{eq:h3-definition}
over one blip or one sojourn interval, hence proportional to 
\begin{eqnarray}
\hbox{Expr}(\Delta t) &=& \int_{t_i}^{t_{i+1}}ds\int_{t_i}^{s} ds' \sum_{b\in C}\frac{C_b^2}{2m_b}\cos\omega_b(s-s') \nonumber \\
&=&  \sum_{b\in C}\frac{C_b^2}{2m_b\omega_b^2} \left(1-\cos\omega_b\Delta t \right)
\end{eqnarray}
For an Ohmic bath with sharp cut-off this expression is $\frac{2\eta}{\pi}\left(\Omega-\delta_{\Omega}(\Delta t)\right)$
where $\delta_{\Omega}(\Delta t)$ a delta-function smoothened at time scale $\Omega^{-1}$.
The contribution to $G_{if}(\nu)$ from $n+1$ sojourns and $n$ blips is hence
\begin{eqnarray}
\hbox{Expr} &=& \frac{2\eta}{\pi} \Omega -\frac{2\eta}{\pi} \Omega + \frac{2\eta}{\pi} \Omega \ldots \nonumber \\
&&- \frac{2\eta}{\pi} \delta_{\Omega}(t_1-t_0)+ \frac{2\eta}{\pi} \delta_{\Omega}(t_2-t_1) \ldots 
\end{eqnarray}
While the first line sums to a large number it does not scale with the time, and there will hence not
be any contribution to thermal power from these terms.

The large terms are in fact an artifact from assuming that the baths are in equilibrium at the start and the end
of the process while still interacting strongly with the system. It has been known for quite some
time that this 
leads to problems already for the open quantum system state~\cite{Grabert1984,Ford1985,daCosta2000,Ingold2009}.
One way to resolve the problem for heat is to assume that the interaction coefficients
$C_b$ depend on time, and vanish in the beginning of the process~\cite{Aurell2017}.
Assuming as in~\cite{Aurell2017} and in analogy with \eqref{eq:C-cL} above that $C_{b}(s)=\sqrt{\omega_c^{3}m_{\omega}\eta(s)}$
we have instead of above
\begin{eqnarray}
\hbox{Expr} &=& \sum_i (-1)^i \left(\frac{1}{4}(\dot{\eta}(t_{i+1})-\dot{\eta}(t_{i}))+ \frac{1}{4}\int_{t_i}^{t_{i+1}} \frac{(\dot{\eta})^2}{\eta} ds
\right)
\end{eqnarray}
In above the bath cut-off frequency has been taken to infinity.
Clearly if the function $\eta(s)$ is constant except at the boundaries this does not give anything
proportional to the duration of the process.

\section{The non-singular NIBA heat terms: general formalism}
\label{sec:non-singular-heat-NIBA}
The main idea 
is to write the sum $G_i(\nu)=\sum_f G_{if}(\nu)$ 
as a matrix product (transfer matrix formalism).
The formulation is as follows:

\begin{enumerate}
\item Starting state $i$ is by convention ``up''.
The starting vector is therefore $\chi_0 =\left(\begin{array}{c}1\\0\end{array}\right)=(\uparrow,\uparrow)$.

\item End vector, when we sum over the final state of the system, is 
$\chi_n = \left(\begin{array}{c}1\\1\end{array}\right)=(\uparrow,\uparrow)+(\downarrow,\downarrow)$.

\item The phase terms at the jumps are determined by the translation tables

      \begin{tabular}{|lr|lr|c|r|} \hline
      \multicolumn{4}{c}{\em sojourn $\to$ blip} \\ \hline
      {\em start state}        & $\chi$     & {\em end state}       &$\xi$ & {\em forward/backward} & {\em factor} \\ \hline
      $\uparrow,\uparrow$      &  +1        & $\uparrow,\downarrow$ &+1    & B                        & $-i\frac{\Delta}{2}$ \\ 
      $\uparrow,\uparrow$      &  +1        & $\downarrow,\uparrow$ &-1    & F                        & $i\frac{\Delta}{2}$ \\ 
      $\downarrow,\downarrow$  &  -1        & $\uparrow,\downarrow$ &+1    & F                        & $i\frac{\Delta}{2}$ \\ 
      $\downarrow,\downarrow$  &  -1        & $\downarrow,\uparrow$ &-1    & B                        & $-i\frac{\Delta}{2}$ \\ \hline
      \end{tabular}

      \vspace{0.4cm}

      and 

      \vspace{0.4cm}
      
      \begin{tabular}{|lr|lr|c|r|} \hline
      \multicolumn{4}{c}{\em blip $\to$ sojourn} \\ \hline
      {\em start state}     &$\xi$   & {\em end state}         &$\chi$    &          {\em forward/backward} & {\em factor} \\ \hline
      $\uparrow,\downarrow$ &+1      & $\uparrow,\uparrow$     &+1        &                               B & $-i\frac{\Delta}{2}$ \\ 
      $\uparrow,\downarrow$ &+1      & $\downarrow,\downarrow$ &-1        &                               F & $i\frac{\Delta}{2}$ \\ 
      $\downarrow,\uparrow$ &-1      & $\uparrow,\uparrow$     &+1        &                               F & $i\frac{\Delta}{2}$ \\ 
      $\downarrow,\uparrow$ &-1      & $\downarrow,\downarrow$ &-1        &                               B & $-i\frac{\Delta}{2}$ \\ \hline
      \end{tabular}    

\item To every transition \textit{sojourn $\to$ blip} are associated terms $e^{\frac{i}{\hbar}\chi_{j-1}\xi_j \left(X_{j.j-1} + \frac{1}{2}\nu K\right)}$.
Combine this and the phase factor to a matrix $\frac{i\Delta}{2} \mathbf{T}$.

\item To every \textit{blip interval} is associated the terms $e^{-\frac{1}{\hbar}S_j-\frac{i}{\hbar}\epsilon (t_{2j}-t_{2j-1})}$.
Call this diagonal matrix $\mathbf{\Lambda}$.

\item To every transition \textit{blip $\to$ sojourn} is associated a term $e^{-\frac{i}{\hbar}\chi_{j}\xi_j \nu\frac{1}{2}K}$ 
Combine this and the phase factors to a matrix $\frac{i\Delta}{2}\mathbf{S}$.

\item To every transition \textit{sojourn $\to$ sojourn} is additionally associated as term
$e^{\frac{i}{\hbar}\chi_{j}\chi_{j-1} \nu\frac{1}{2}\tilde{K}}$. This is the same for both signs of the blip in between. 

\item The transition \textit{sojourn $\to$ sojourn} is then given by a matrix $\mathbf{M}$
formed by $\mathbf{S}\mathbf{\Lambda}\mathbf{T}$ and the modifications due to $\tilde{K}$.
By matrix multiplication one finds
      \begin{equation}
      \mathbf{M} = e^{-\frac{1}{\hbar} S}\left(\begin{array}{ll} 
                2\cos\frac{1}{\hbar}(X-\epsilon t) e^{\frac{i}{\hbar}\frac{1}{2}\nu\tilde{K}}& 
               -2\cos\frac{1}{\hbar}(X+\nu K+\epsilon  t) e^{-\frac{i}{\hbar}\frac{1}{2}\nu\tilde{K}}\\
               -2\cos\frac{1}{\hbar}(X+\nu K-\epsilon t) e^{-\frac{i}{\hbar}\frac{1}{2}\nu\tilde{K}}& 
                2\cos\frac{1}{\hbar}(X+\epsilon t)e^{\frac{i}{\hbar}\frac{1}{2}\nu\tilde{K}}
                                                        \end{array}\right)
       \end{equation}
       For simplicity the blip interval is written $t$.

\item The whole generating function can hence, within NIBA, be written as

      \begin{equation}
      \label{eq:generating-function-sum}
      G_{i}(\nu) = \left(\begin{array}{ll}1 & 1\end{array}\right)\left(\sum_n (-1)^n (\frac{\Delta}{2})^{2n} \mathbf{M}^n\right)  \left(\begin{array}{c}1\\0\end{array}\right)
      \end{equation}

      where all the blip times are implicit in the matrices $\mathbf{M}$ on the right-hand side.
\end{enumerate}
To analyze \eqref{eq:generating-function-sum} in a stationary setting (the bias $\epsilon$ and
all other parameters are constant in time) one takes a Laplace transform.
Every sojourn interval then yields a factor $\lambda^{-1}$, and the $n$'th term in \eqref{eq:generating-function-sum} hence a factor $\lambda^{-1-n}$.
For the Laplace transform of the matrix it is convenient to write
\begin{equation}
      \tilde{\mathbf{M}}(\lambda) = 2\left(\begin{array}{ll} A(\lambda) & -B(\lambda,\nu) \\                                                            
                                                            -C(\lambda,\nu) & D(\lambda)\end{array}\right)
\end{equation}
where 
\begin{eqnarray}
\label{eq:A-def}
A &=& \int d t e^{-\lambda t} e^{-\frac{1}{\hbar} S} \cos\frac{1}{\hbar}(X-\epsilon t) e^{\frac{i}{\hbar}\frac{1}{2} \nu\tilde{K}}\\
\label{eq:B-def}
B &=& \int d t e^{-\lambda t} e^{-\frac{1}{\hbar} S} \cos\frac{1}{\hbar}(X+\nu K+\epsilon t)e^{-\frac{i}{\hbar}\frac{1}{2}\nu\tilde{K}}\\
\label{eq:C-def}
C &=& \int d t e^{-\lambda t} e^{-\frac{1}{\hbar} S} \cos\frac{1}{\hbar}(X+\nu K-\epsilon t)e^{-\frac{i}{\hbar}\frac{1}{2}\nu\tilde{K}} \\
\label{eq:D-def}
D &=& \int d t e^{-\lambda t} e^{-\frac{1}{\hbar} S} \cos\frac{1}{\hbar}(X+\epsilon t)e^{\frac{i}{\hbar}\frac{1}{2}\nu\tilde{K}}
\end{eqnarray}
All $S$, $X$, $K$ and $\tilde{K}$ depend on the blip time $t$ (at least in principle).

The Laplace transform of the generating function is
\begin{eqnarray}
       \hat{G}_{i}(\nu,\lambda) &=& \int d t e^{-\lambda t} G_{i}(\nu,t) \nonumber \\ 
&=& \lambda^{-1}\left(\begin{array}{ll}1 & 1\end{array}\right)\left(\sum_n (-1)^n\lambda^{-n}(\frac{\Delta}{2})^{2n} \tilde{\mathbf{M}}^n\right) \left(\begin{array}{c}1\\0\end{array}\right)
\label{eq:generating-function-formula-heat-sum}
\end{eqnarray}

\section{The generating function at $\nu=0$}
\label{sec:generating-function-value}
The special case of $\nu=0$ is an important check, because that should give the quantity 
computed by Leggett in~\cite{Leggett87}:
       $\tilde{P}(\lambda) = \int d t e^{-\lambda t} \left<\sigma_z\right>(t)$.
The relation is $\left<\sigma_z\right>(t)=2\cdot\hbox{Prob}(\hbox{"up"},t)-1$
and hence $\tilde{P}(\lambda) = \tilde{G}_{if}(\nu=0,\lambda) -\lambda^{-1}$
where $i$ and $f$ are both ``up''. The formula found by Leggett is
\begin{equation}
\label{eq:Leggett-7.6}
\tilde{P}(\lambda) = \frac{1-\tilde{h}/\lambda}{\lambda+\tilde{g}}\quad(\hbox{\cite{Leggett87}, Eq.~7.6})
\end{equation}
where 
\begin{eqnarray}
\label{eq:Leggett-7.5a}
\tilde{g} &=& \int d t e^{-\lambda t} \Delta^2 e^{-\frac{1}{\hbar}S}\cos\frac{1}{\hbar}X\cos\frac{\epsilon t}{\hbar}
\quad(\hbox{\cite{Leggett87}, Eq.~7.5a}) \\
\label{eq:Leggett-7.5b}
\tilde{h} &=& \int d t e^{-\lambda t} \Delta^2 e^{-\frac{1}{\hbar}S}\sin\frac{1}{\hbar}X\sin\frac{\epsilon t}{\hbar}
\quad(\hbox{\cite{Leggett87}, Eq.~7.5b}) 
\end{eqnarray}

We hence consider \eqref{eq:generating-function-formula-heat-sum} at $\nu=0$.
We have the simplification that $C=A$ and $B=D$, and the Laplace transform matrix is hence
\begin{equation}
      \tilde{\mathbf{M}}(\lambda) = 2\left(\begin{array}{ll} A & -D \\                                                            
                                                            -A & D\end{array}\right)
\end{equation}
The eigenvalues of this matrix are $0$ and $2(A+D)$. Positive powers of this matrix ($n\geq 1$) are thus simply
\begin{equation}
      \left(\tilde{\mathbf{M}}(\lambda)\right)^n = \left(2(A+D)\right)^{n-1}\tilde{\mathbf{M}}(\lambda)
\end{equation}
which means that 
\begin{eqnarray}
\label{eq:G-if-lambda-A-and-D}
    G_{if}(\nu=0,\lambda) &=& \lambda^{-1} - \lambda^{-2}\frac{(\Delta}{2})^2\frac{2A}{1+\lambda^{-1}\frac{\Delta^2}{2}(A+D)}
\end{eqnarray}
We may identify $\frac{\Delta^2}{2}(A+D)=\tilde{g}$ and 
$\frac{\Delta^2}{2}A=\frac{1}{2}(\tilde{g}+\tilde{h})$ and so
\begin{eqnarray}
\label{eq:G-if-lambda-g-and-h}
    G_{if}(\nu=0,\lambda) &=& \lambda^{-1} - \lambda^{-2}\frac{1}{2}\frac{\tilde{g}+\tilde{h}}{1+\lambda^{-1}\tilde{g}}
\end{eqnarray} 
This means that
\begin{eqnarray}
    \tilde{P} &=& \lambda^{-1} - \lambda^{-2}\frac{\tilde{g}+\tilde{h}}{1+\lambda^{-1}\tilde{g}}
=  \lambda^{-1}\frac{\lambda +\tilde{g}-\tilde{g}-\tilde{h}}{\lambda+\tilde{g}} 
\end{eqnarray} 
which is \eqref{eq:Leggett-7.6}, as required. The result $G_{i}(\nu=0,t)=1$ (normalization of the system state)
follows from 
$\left(\begin{array}{ll}1 & 1\end{array}\right)\tilde{\mathbf{M}}=0$,
which means that ${G}_{i}(\nu=0,\lambda)=\lambda^{-1}$ (only $n=0$ term survives).

\section{The long term limit of the generating function at $\nu=0$}
\label{sec:long-term-limit}
On physical grounds it is reasonable to assume that for long times
the generating function is 
\begin{equation}
  G_{if}(\nu,t)_{\nu=0}= p + \sum_k q_k e^{-t r_k} 
\end{equation}
where $p$ is the long term limit of the probability to be up, and $q_k$ and $r_k$ are some constants.
The Laplace transform is then
\begin{equation}
  \hat{G}_{if}(\nu,\lambda)_{\nu=0}= p\lambda^{-1} + \sum_k \frac{q_k}{\lambda+ r_k} 
\end{equation}
from which follows
\begin{equation}
  p=\lim_{\lambda \rightarrow 0} \lambda \tilde{G}_{i}(\nu,\lambda)_{\nu=0}
\end{equation}
Inserting \eqref{eq:G-if-lambda-A-and-D}
we have 
\begin{equation}
  \label{eq:p-result}
  p= \frac{D}{A+D}
\end{equation}
where in the integrals defining $A$ and $D$ the Laplace transform parameter $\lambda$ is zero.

A physical density matrix of the qubit must lie inside the Bloch sphere.
A necessary condition for  
$\frac{D}{A+D}$ and $\frac{A}{A+D}$ to be the diagonal elements
of a stationary density matrix in the long-time limit 
is hence that they fall between zero and one.
For a qubit interacting with one bath at one temperature 
this was shown to be always the case in~\cite{Leggett87},
even when the density matrix computed under these assumption of NIBA is not correct.

For our case of one qubit interacting with two baths the situation
is more involved, and we state it as
\begin{theorem}
Consider $S=S_C+S_H$ and $X=X_C+X_H$ as an even and an odd function on the whole line.
Let $\hat{F}(\omega)$ be the Fourier transform of
$e^{-\frac{1}{\hbar}S+\frac{i}{\hbar}X}$
and $\hat{F}^*(\omega)=\hat{F}(-\omega)$ the Fourier transform of
$e^{-\frac{1}{\hbar}S-\frac{i}{\hbar}X}$.
Then $\frac{D}{A+D}$ and $\frac{A}{A+D}$ 
are possible diagonal elements of a density matrix
if $|{\cal I}\hat{F}(\frac{\epsilon}{\hbar})|<|{\cal R}\hat{F}(\frac{\epsilon}{\hbar})|$.
\end{theorem}
The proof is by simple translation.
We may write
\begin{equation}
\frac{A}{A+D} = \frac{1}{2} + \frac{1}{2}
   \frac{\int d t  e^{-\frac{1}{\hbar} S} \sin\frac{1}{\hbar}X\sin\frac{1}{\hbar}\epsilon t}
        {\int d t  e^{-\frac{1}{\hbar} S} \cos\frac{1}{\hbar}X\cos\frac{1}{\hbar}\epsilon t}
\end{equation}
and the condition
\begin{equation}
0\leq \frac{A}{A+D}\leq 1
\end{equation}
is hence the same as
\begin{equation}
\left|\int d t  e^{-\frac{1}{\hbar} S} \sin\frac{1}{\hbar}X\sin\frac{1}{\hbar}\epsilon t\right|
\leq \left|\int d t  e^{-\frac{1}{\hbar} S} \cos\frac{1}{\hbar}X\cos\frac{1}{\hbar}\epsilon t\right|
\end{equation}
Multiplying out and identifying terms says that the imaginary part of the Fourier transform should be smaller in absolute value than the real part, at the frequency of the level splitting. 
Note that the theorem does not give a condition for NIBA with two baths
to be correct, only a condition for it to give physically admissible populations.

With two caveats one may interpret~\eqref{eq:p-result} 
in an almost classical manner.
First we can (trivially) rewrite it as
\begin{equation}
  \label{eq:p-result-2}
  p= \frac{\frac{\Delta^2}{2}D}{\frac{\Delta^2}{2}A+\frac{\Delta^2}{2}D}
\end{equation}
where (at $\lambda=0$) 
\begin{eqnarray}
\label{eq:A-def-2}
\frac{\Delta^2}{2}A &=& \left(i\frac{\Delta}{2}\right) \left(-i\frac{\Delta}{2}\right) 
\int d t e^{-\frac{1}{\hbar} S} \left(e^{\frac{i}{\hbar}(X-\epsilon t)}+e^{\frac{i}{\hbar}(-X+\epsilon t)}\right)\\
\label{eq:D-def-2}
\frac{\Delta^2}{2}D &=& \left(i\frac{\Delta}{2}\right) \left(-i\frac{\Delta}{2}\right) 
\int d t e^{-\frac{1}{\hbar} S} 
\left(e^{\frac{i}{\hbar}(X+\epsilon t)}+e^{\frac{i}{\hbar}(-X-\epsilon t)}\right)
\end{eqnarray}
The two terms in $\frac{\Delta^2}{2}A$ are the integrals over time $t$
of the influence 
functionals of two particular spin histories,
where the state is $(\uparrow,\uparrow)$ before time zero,
then at time zero either the forward or the backward path jumps 
to down, and then at time $t$ the other path follows.
The two terms $\left(i\frac{\Delta}{2}\right)$ and $\left(-i\frac{\Delta}{2}\right)$
are the jump rate amplitudes (dimension $\left(\hbox{time}\right)^{-1}$) for the two paths. 
These combined with the integral over time $t$ 
hence gives a quantitity analogous to 
the probability that the state transits from $(\uparrow,\uparrow)$ to  $(\downarrow,\downarrow)$
per unit time. 
The two terms in $\frac{\Delta^2}{2}D$ 
may similarly be taken to represent 
the total rate of the state transiting from $(\downarrow,\downarrow)$
to $(\uparrow,\uparrow)$.

The first of the two caveat is that by the above 
$A$ and $D$ may have different signs so that one of 
$\frac{A}{A+D}$ and $\frac{D}{A+D}$ is negative, and
the other is larger than one. If so, NIBA would not 
give a physically admissable state.
The second is that even when $\frac{A}{A+D}$ and $\frac{D}{A+D}$    
are both between zero and one, both $A$ and $D$ could be negative.
NIBA would in that case give a physically admissable state,
but not one that can be described as from a classical jump process.

%The seemingly complicated quantum result 
%\eqref{eq:p-result-2} is thus actually very analogous to 
%the stationary probability in a simple classical jump process
%with rates $k_{\rightarrow}$ (up to down)  and $k_{\leftarrow}$ (down to up): 
%\begin{equation}
%  \label{eq:p-result-3}
%  \hbox{Prob}\left(\uparrow\right) = \frac{k_{\leftarrow}}{k_{\leftarrow}+k_{\rightarrow}}
%\end{equation}

\section{Derivatives of generating function formula at $\nu=0$}
\label{sec:generating-function-derivatives}
The expected energy change of the bath is given by the derivative of the generating
function \eqref{eq:generating-function-formula-heat-sum} with respect to $i\nu$ taken at $\nu=0$.
At any $\nu$ this quantity is
\begin{eqnarray}
       \frac{d}{d(i\nu)}\hat{G}_{i}(\nu,\lambda) 
&=& -\lambda^{-2}(\frac{\Delta}{2})^2 \left(\begin{array}{ll}1 & 1\end{array}\right)
\left(\sum_l (-1)^l\lambda^{-l}(\frac{\Delta}{2})^{2l}\tilde{\mathbf{M}}^l\right)\frac{d \tilde{\mathbf{M}}}{d(i\nu)}
\left(\sum_k (-1)^k\lambda^{-k}(\frac{\Delta}{2})^{2k}\tilde{\mathbf{M}}^k\right)\left(\begin{array}{c}1\\0\end{array}\right)
\label{eq:generating-function-formula-heat-derivative}
\end{eqnarray}

At $\nu=0$ the sums on the left and the right simplify as above.
\textit{On the left} only the zeroth order term ($l=0$) survives,
while \textit{on the right} we have
\begin{equation}
\left(\sum_k (-1)^k\lambda^{-k}(\frac{\Delta}{2})^{2k}\tilde{\mathbf{M}}^k\right)\left(\begin{array}{c}1\\0\end{array}\right)
= \left(\begin{array}{c}1\\0\end{array}\right) - \lambda^{-1}(\frac{\Delta}{2})^{2} \frac{2A}{1+\lambda^{-1}\frac{\Delta^2}{2}(A+D)}
 \left(\begin{array}{c}1\\-1\end{array}\right)
\end{equation}

The dependence on $\nu$ comes either through the function $K$, or the function $\tilde{K}$.
In the first case only the off-diagonal elements ($B$ and $C$) depend on $\nu$,
and the total expression is
\begin{eqnarray}
       \frac{d}{d(i\nu)}\hat{G}_{i}(\nu,\lambda)|_{\nu=0,\, \hbox{through $K$}} 
&=& \lambda^{-2}(\frac{\Delta}{2})^2 2\dot{C} \nonumber \\
&& - \lambda^{-3}(\frac{\Delta}{2})^4 \frac{2A}{1+\lambda^{-1}\frac{\Delta^2}{2}(A+D)} 2(\dot{C}-\dot{B})   
\label{eq:generating-function-formula-heat-derivative-2}
\end{eqnarray}
where $\dot{C}=\frac{dC}{d(i\nu)}|_{\nu=0\, \hbox{through $K$}}$ and $\dot{B}=\frac{dB}{d(i\nu)}|_{\nu=0\, \hbox{through $K$}}$.
These derivatives follow from \eqref{eq:B-def} and  \eqref{eq:C-def} and are
\begin{eqnarray}  
\label{eq:dot-B-def}
\dot{B} &=& \int d t e^{-\lambda t} e^{-\frac{1}{\hbar} S} \sin\frac{1}{\hbar}(X+\epsilon t)\left(\frac{i}{\hbar}K\right)  \\
\label{eq:dot-C-def}
\dot{C} &=& \int d t  e^{-\lambda t} e^{-\frac{1}{\hbar} S} \sin\frac{1}{\hbar}(X-\epsilon t)\left(\frac{i}{\hbar}K\right) 
\end{eqnarray}
Following \eqref{eq:K-and-S} we can rewrite this as
\begin{eqnarray}  
\label{eq:dot-B-def-v2}
\dot{B} &=& \int d t e^{-\lambda t} e^{-\frac{1}{\hbar} S} \sin\frac{1}{\hbar}(X+\epsilon t)\left(-\frac{dS}{dt}\right)  \\
\label{eq:dot-C-def-v2}
\dot{C} &=& \int d t  e^{-\lambda t} e^{-\frac{1}{\hbar} S} \sin\frac{1}{\hbar}(X-\epsilon t)\left(-\frac{dS}{dt}\right) 
\end{eqnarray}

In the second case of dependence through 
$\tilde{K}$ the derivative matrix is
      \begin{equation}
      \frac{d\mathbf{M}}{d(i\nu)}|_{\nu=0,\, \hbox{through $\tilde{K}$}} = e^{-\frac{1}{\hbar} S} \left(\begin{array}{ll} 
                \cos\frac{1}{\hbar}(X-\epsilon t) \frac{dX}{dt}& 
                \cos\frac{1}{\hbar}(X+\epsilon  t)\frac{dX}{dt} \\
                \cos\frac{1}{\hbar}(X-\epsilon t) \frac{dX}{dt}& 
                \cos\frac{1}{\hbar}(X+\epsilon t) \frac{dX}{dt}
                                                        \end{array}\right)
       \end{equation}
where we have used \eqref{eq:tildeK-and-X}.
Together with \eqref{eq:generating-function-formula-heat-derivative-2} we have hence also
\begin{eqnarray}
       \frac{d}{d(i\nu)}\hat{G}_{i}(\nu,\lambda)|_{\nu=0,\, \hbox{through $\tilde{K}$}} 
&=& -\lambda^{-2}(\frac{\Delta}{2})^2 2 A' \nonumber \\
&& + \lambda^{-3}(\frac{\Delta}{2})^4 \frac{2A}{1+\lambda^{-1}\frac{\Delta^2}{2}(A+D)} 2(A'-D')   
\label{eq:generating-function-formula-heat-derivative-3}
\end{eqnarray}
where
\begin{eqnarray}  
\label{eq:A-prime}
A' &=& \int d t e^{-\lambda t} e^{-\frac{1}{\hbar} S} \cos\frac{1}{\hbar}(X-\epsilon t)\left(\frac{dX}{dt}\right)  \\
\label{eq:D-prime}
D' &=& \int d t  e^{-\lambda t} e^{-\frac{1}{\hbar} S} \cos\frac{1}{\hbar}(X+\epsilon t)\left(\frac{dX}{dt}\right) 
\end{eqnarray}

\section{Long-time limit of the derivative}
\label{sec:long-term-limit-derivative}
On physical grounds it is reasonable to assume that
the derivative of the generating function with respect to its argument is for long times
\begin{equation}
  \frac{d}{d(i\nu)}G_{i}(\nu,t)_{\nu=0}= \Pi\cdot t + b + \sum_k c_k e^{-t\lambda_k} 
\end{equation}
where $\Pi$ is the long time limit of the power (heat per unit time), and $b$, $c_k$ and $\lambda_k$ are some constants.
The Laplace transform is then
\begin{equation}
  \frac{d}{d(i\nu)}\tilde{G}_{i}(\nu,\lambda)_{\nu=0}= \Pi\lambda^{-2} + b\lambda^{-1} + \sum_k \frac{c_k}{\lambda+ \lambda_k} 
\end{equation}
from which follows
\begin{equation}
  \Pi=\lim_{\lambda \rightarrow 0} \lambda^2\frac{d}{d(i\nu)}\tilde{G}_{i}(\nu,\lambda)_{\nu=0}
\end{equation}
Inserting the various formulas above we have 
\begin{eqnarray}
  \label{eq:a-result}
  \Pi&=& \frac{D}{A+D}\frac{\Delta^2}{2}\left(\int dt e^{-\frac{1}{\hbar}S}\sin\frac{1}{\hbar}(X-\epsilon t)(-\partial_t S_C)
                                       + \int dt e^{-\frac{1}{\hbar}S}\cos\frac{1}{\hbar}(X-\epsilon t)(\partial_t X_C)\right) \nonumber \\
  &+& \frac{A}{A+D}\frac{\Delta^2}{2}\left(\int dt e^{-\frac{1}{\hbar}S}\sin\frac{1}{\hbar}(X+\epsilon t)(-\partial_t S_C)
                                       + \int dt e^{-\frac{1}{\hbar}S}\cos\frac{1}{\hbar}(X+\epsilon t)(\partial_t X_C)\right) 
\end{eqnarray}
where in the integrals defining $A$ and $D$ the Laplace transform parameter $\lambda$ is zero,
and where the subscript $C$ indicates that only the quantities for the cold bath are considered.
Clearly we now have an expression for power similar to the dimensional formula~\eqref{eq:dimensional-argument}.
For the case of only one bath we can integrate by parts in \eqref{eq:a-result} to get
\begin{eqnarray}
  \label{eq:a-result-one-bath}
  \hbox{One bath: } \Pi&=& \frac{D}{A+D}\frac{\Delta^2}{2}\left(
                                       (\epsilon)\int dt e^{-\frac{1}{\hbar}S}\cos\frac{1}{\hbar}(X-\epsilon t)\right) 
  + \frac{A}{A+D}\frac{\Delta^2}{2}\left(
                                       (-\epsilon)\int dt e^{-\frac{1}{\hbar}S}\cos\frac{1}{\hbar}(X+\epsilon t)\right) \nonumber \\
  &=& \frac{D}{A+D}\frac{\Delta^2}{2}(\epsilon A) + \frac{A}{A+D}\frac{\Delta^2}{2}(-\epsilon D) = 0
\end{eqnarray}
which is the expected result. In the long term limit the thermal power from one qubit equilibrating with one bath must vanish.
If we were to consider heat to the hot bath, all that would change 
\eqref{eq:a-result} is that the time derivatives would be $\partial_t S_H$
and $\partial_t X_H$. By adding the same argument 
as in  \eqref{eq:a-result-one-bath} shows that the
the sum of thermal power to the cold bath and the hot
bath cancel.

In the case of two baths 
and heat to one bath it is on the other hand more convenient to write
$S=S_C+S_H$ and $X=X_C+X_H$ and to introduce the kernels~\footnote{Similar kernels have been introduced in the literature
before, but not exactly these ones; hence the new notation.}
\begin{eqnarray}
C_+^C(t) &=& e^{-\frac{1}{\hbar}S_C + \frac{i}{\hbar}X_C} \\ 
C_+^H(t) &=& e^{-\frac{1}{\hbar}S_H + \frac{i}{\hbar}X_H} \\ 
C_-^C(t) &=& e^{-\frac{1}{\hbar}S_C - \frac{i}{\hbar}X_C} \\ 
C_-^H(t) &=& e^{-\frac{1}{\hbar}S_H - \frac{i}{\hbar}X_H}  
\end{eqnarray}
in terms of which \eqref{eq:a-result} can be written
\begin{eqnarray}
  \label{eq:a-result-two-baths}
  \hbox{Two baths: } \Pi&=& \frac{D}{A+D}\frac{\Delta^2}{4} \left( -i\hbar\int dt e^{-\frac{i\epsilon t}{\hbar}}\frac{dC_+^C(t)}{dt} C_+^H(t)  
                       +i\hbar \int dt e^{\frac{i\epsilon t}{\hbar}}\frac{dC_-^C(t)}{dt} C_-^H(t) \right) \nonumber \\ 
                      &&+ \frac{A}{A+D}\frac{\Delta^2}{4}\left( -i\hbar\int dt e^{\frac{i\epsilon t}{\hbar}}\frac{dC_+^C(t)}{dt} C_+^H(t)  
                       +i\hbar \int dt e^{\frac{-i\epsilon t}{\hbar}}\frac{dC_-^C(t)}{dt} C_-^H(t) \right) 
\end{eqnarray}
This is the formulation used in Section~\ref{sec:strong-coupling} and Section~\ref{sec:polaron-transform} in the main text.

Physically, thermal power to the cold bath must be positive.
Referring to the discussion at the end of Appendix~\ref{sec:long-term-limit}
we may identify $A$ as $\frac{1}{2}\hat{F}^{*}(\frac{\epsilon}{\hbar})$
and $D$ as $\frac{1}{2}\hat{F}(\frac{\epsilon}{\hbar})$
and the terms 
in parentheses in
\eqref{eq:a-result-two-baths}
as Fourier components of the function $H(t)=i\hbar \frac{dC_+^C(t)}{dt} C_+^H(t)$.
Thermal power would then be 
$-\frac{\Delta^2}{2{\cal R}[\hat{F}(\frac{\epsilon}{\hbar})]}
{\cal R}[\hat{F}(\frac{\epsilon}{\hbar})\hat{H}^*(\frac{\epsilon}{\hbar})]$.

\end{widetext}

\end{document}